# Carbon substitution in MgB$_2$ single crystals: structural and superconducting properties

S. M. Kazakov,[1] R. Puzniak,[1,2] K. Rogacki,[1,3] A. V. Mironov,[4] N. D. Zhigadlo,[1] J. Jun,[1] Ch. Soltmann,[1] B. Batlogg,[1] and J. Karpinski[1]

[1] Solid State Physics Laboratory ETH, 8093 Zürich, Switzerland

[2] Institute of Physics, Polish Academy of Sciences, Al. Lotników 32/46, PL 02-668 Warszawa, Poland

[3] Institute of Low Temperature and Structure Research, Polish Academy of Sciences, P.O. Box 1410, 50-950 Wroclaw, Poland

[4] Moscow State University, 119899 Moscow, Russia

Growth of carbon substituted magnesium diboride Mg(B$_{1-x}$C$_x$)$_2$ single crystals with 0≤x≤0.15 is reported and the structural, transport, and magnetization data are presented. The superconducting transition temperature decreases monotonically with increasing carbon content in the full investigated range of substitution. By adjusting the nominal composition, $T_c$ of substituted crystals can be tuned in a wide temperature range between 10 and 39 K. Simultaneous introduction of disorder by carbon substitution and significant increase of the upper critical field $H_{c2}$ is observed. Comparing with the non-substituted compound, $H_{c2}$ at 15K for x=0.05 is enhanced by more then a factor of 2 for $H$ oriented both perpendicular and parallel to the $ab$-plane. This enhancement is accompanied by a reduction of the $H_{c2}$-anisotropy coefficient $\gamma$ from 4.5 (for non-substituted compound) to 3.4 and 2.8 for the crystals with $x$ = 0.05 and 0.095, respectively. At temperatures below 10 K, the single crystal with larger carbon content shows $H_{c2}$ (defined at zero resistance) higher than 7 and 24 T for $H$ oriented perpendicular and parallel to the $ab$-plane, respectively. Observed increase of $H_{c2}$ cannot be explained by the change in the coherence length due to disorder-induced decrease of the mean free path only.

# I. INTRODUCTION

For the possible large-scale applications of the $MgB_2$ superconducor the upper critical field $H_{c2}$ is a central materials parameter.[1] For pure single crystals the out of plane $H_{c2}^{//c}(0)$ is as low as 3.1 T (Ref. 2) and therefore, many efforts have been made to modify the superconducting properties of this material through chemical substitution. A brief review of the current status of the substitutional chemistry of $MgB_2$ can be found in the paper by Cava *et al.*[3] Carbon substitution for boron appeared to be one of the most interesting, though the reported carbon solubility in $MgB_2$ and its influence on $T_c$ varies considerably depending on the synthesis route and starting materials.[4, 5, 6, 7] Carbon solubility ranging from 1.25% to 30% was reported[4, 6] when elemental magnesium, boron and carbon were used as starting materials. Another approach for carbon substitution of $MgB_2$ was suggested by Mickelson *et al.*[8] Using boron carbide ($B_4C$) as a source of carbon they achieved better mixing of B and C atoms. The samples had an estimated composition $Mg(B_{0.9}C_{0.1})_2$ and $T_c$ was decreased by 7 K. Ribeiro *et al.* followed this approach and reported the optimization of synthetic conditions of carbon doped samples leading to nearly single phase $Mg(B_{1-x}C_x)_2$ with $T_c$ = 22 K.[9] Neutron diffraction study performed on their sample yielded a carbon substitution level of about 10%, and a linear relation between the unit cell parameter *a* and the C concentration was found.[10] The authors showed that earlier reports overestimated the carbon content in $Mg(B_{1-x}C_x)_2$ because polycrystalline carbon substituted samples may contain significant amounts of impurity phases and the nominal content was assumed most often to be equal to actual one. Refinement of neutron diffraction data performed on very carefully prepared polycrystalline samples showed that samples contain ~73% of the $Mg(B_{1-x}C_x)_2$ phase.[10]

The improvement of flux pinning and the enhancement of the critical current density in $MgB_2$ powder by both nanoparticle SiC (Ref. 11) and nano-carbon doping[12] was reported by Dou *et al.* They claimed that simultaneous Si and C substitution for B raised the saturation limit of C in $MgB_2$ considerably while, the $T_c$ reduction is not pronounced (only 2.4 K for $x$ = 0.4). The shift of the irreversibility line towards higher fields and temperatures, caused by nano-diamond doping to $MgB_2$, was reported by Cheng *et al.*[13]

Most of these substitutions involved polycrystalline samples. The availability of single crystals is expected to provide more detailed insight into the microscopic aspects of substitution, including solubility limit, and $T_c$ variation as a function of doping as well as the impact of disorder-induced defects on superconducting properties. Recently, several reports on carbon-substituted single crystals appeared: one of them presents shortly our results of the $T_c$ dependence on carbon content[14]; Lee *et al.* presented lattice parameters, magnetization and resistivity measurement data for $MgB_2$ crystals with carbon content up to 15% (Ref. 15). The data of the last report are in qualitative agreement with ours, presented here, however quantitative differences exists. Lee *et al.* presented nominal carbon content $x$ in all of their plots and assumed that it is equal to the actual one, however, according to our experience the nominal carbon content for single crystals of $Mg(B_{1-x}C_x)_2$ is different from the one calculated from the *a* lattice parameter dependence.

In the context of recent discussions on the transport properties of $MgB_2$ (Ref. 16), it is of general interest to study the temperature dependence of the resistivity of single phase compounds, where both the mean free path and the superconducting coherence length can vary considerably due to substitutions. Single crystals instead of polycrystalline materials are preferred to separate granularity effects from changes in the electron structure or the electron scattering process. Pure and aluminum substituted $MgB_2$ has been examined widely.[3, 17, 18, 19, 20] However, the carbon-substituted compounds have been studied only recently in both polycrystalline samples[21, 22] and in single crystals.[15, 23, 24, 25] The value and the anisotropy of the upper critical field $H_{c2}$ for the $Mg(B_{1-x}C_x)_2$ compounds have been reported in a limited range.[23, 24, 25] At 15 K, $H_{c2}$ for optimally C-substituted $MgB_2$ single crystals show an increase by a factor of 2 for $H$ applied parallel as well as perpendicular to the $ab$-plane, however this result depends somewhat on the $H_{c2}$ definition.[26, 27, 28]

In this paper, we report the results of single crystal growth, structure, resistivity and magnetic property investigations of carbon substituted $MgB_2$. We describe $H_{c2}$ deduces from resistance measurements for both field configurations, for $Mg(B_{1-x}C_x)_2$ crystals with $x = 0.05$ and 0.095. Here, the critical temperature $T_c$ at constant magnetic field is defined as a temperature where the non-zero resistant state begins to develop in the zero-current limit. Introduction of disorder by carbon substitution increases $H_{c2}$ as compared to that of $MgB_2$. However, the twofold increase of $H_{c2}$ combined with the modest decrease of $T_c$ by about 5 K cannot be explained by changes in the coherence length caused by the defect-induced decrease of the mean free path only.

## II. EXPERIMENTAL

Single crystals of $Mg(B_{1-x}C_x)_2$ were grown under high pressure using the cubic anvil press. The applied pressure/temperature conditions for the growth of $MgB_2$ single crystals were determined in our earlier study of Mg-B-N phase diagram.[29] Magnesium (Fluka, >99% purity), amorphous boron (Alfa Aesar, >99,99%), carbon graphite powder (Alfa Aesar, >99,99%) and silicon carbide (Alfa Aesar, >99.8%) were used as starting materials. Amorphous boron was annealed under dynamic vacuum at 1200 ºC to minimize contamination by oxygen. Two types of precursors were prepared: in the first case, graphite served as a source of carbon; in the second type of precursors, SiC was added to a mixture of Mg and amorphous B. Starting materials with different nominal carbon content were mixed and pressed. A pellet was put into a BN container of 6 mm internal diameter and 8 mm length. Crystals were grown in the same way as the unsubstituted crystals. First, pressure was applied using a pyrophylite pressure transmitting cube as a medium, then the temperature was increased during one hour, up to the maximum of 1900-1950 ºC, kept for 30 min, and decreased over 1-2 hours.

The single crystal x-ray diffraction investigations were carried out using two four-circle x-ray diffractometers (Siemens P4 and CAD4). The refinement of the crystal structure was made with the JANA2000 program package.[30] Electrical resistivity measurements were performed with 14 T Quantum Design physical property measurement system. Small single crystals were selected (typically with dimensions of 0.5 x 0.3 x 0.04 mm$^3$) to reduce any influence of crystal imperfections. The measuring current density was $i = 2.5$ A/cm$^2$, to study the upper critical field properties, and varied from 0.25 to 40 A/cm$^2$ to examine the vortex dynamics, including *I-V* characteristics. The measurement current flowed in the *ab*-plane and was perpendicular to the applied field. The field was oriented both parallel and perpendicular to the *ab*-plane, which is the main plane of the MgB$_2$ single crystal. Magnetic measurements have been performed with a non-commercial SQUID magnetometer. The temperature dependence of the dc magnetization in an external magnetic fields of $H = 1-3$ Oe was recorded for both zero field cooled and field cooled conditions. Torque measurements were performed in a 9 Tesla Quantum Design PPMS.

## III. RESULTS
### A. Crystal growth

Carbon substituted Mg(B$_{1-x}$C$_x$)$_2$ crystals were grown with dimensions up to 0.8×0.8×0.02 mm$^3$. They were black in color in contrast to golden non-substituted MgB$_2$. Table I presents starting composition, estimated carbon content, lattice parameters as well as the superconducting transition temperature of unsubstituted and carbon substituted MgB$_2$ single crystals. Energy dispersive x-ray (EDX) analysis and laser ablation inductive coupled plasma (ICP) mass spectroscopy analyses performed on the SiC-doped crystals show only C and no traces of Si in the crystals. X-ray single crystal analysis confirmed this finding (see next section). The carbon content in the crystals was estimated from the *a* lattice parameter, assuming the linear dependence of the *a* parameter on carbon content, according to Ref. 10. Although it is a rough approximation, no other proof of carbon content was found and the x-ray refinement was not possible because of similar scattering factors for B and C. The variation of $T_c$ as a function of lattice parameter *a* and carbon concentration is presented in Fig. 1(a) and 1(b) together with the temperature dependence of magnetization curves (inset to Fig. 1(b)). The superconducting transitions are relatively sharp even in highly doped crystals. With increasing C content, $T_c$ decreases smoothly, and we note a more effective suppression of $T_c$ when the C level exceeds ~10% (Fig. 1). By adjusting the carbon content in the starting mixture one can tune $T_c$ in a wide range (39-10 K) for carbon content between $x = 0$ and 0.15.

### B. Single crystal structure analysis

Three single crystals, MgB$_2$, Mg(B$_{0.896}$C$_{0.104}$)$_2$ and Mg(B$_{0.850}$C$_{0.150}$)$_2$, were used for detailed structure analysis. The changes in the structures were expected to be minor and therefore all experiments were handled under the same conditions. Pure MgB$_2$ was refined for comparison. Cell parameters for all of the crystals were refined from one and the same set of 23 reflections in a wide range of $\theta$ angle. No superstructure reflections were found, thus we conclude a statistical distribution of carbon or silicon on the B site. Preliminary study of reflection profiles revealed a highly anisotropic broadening for the substituted crystals. The two-dimensional profiles of reflections for Mg(B$_{0.850}$C$_{0.150}$)$_2$ show the elongation of the reflections along the *c\** direction in reciprocal space (Fig. 2(a)), while in the *a\*b\** plane the reflection profiles (Fig. 2(b)) are similar to those of the unsubstituted phase. For Mg(B$_{0.896}$C$_{0.104}$)$_2$ the elongation is smaller. In pure MgB$_2$ the reflection profiles are the same in all direction of reciprocal space. The anisotropy of the reflections in carbon substituted crystals indicates of crystal disorder, but the structural investigations alone can not identify the type of disorder. Microscopic phase separation suggested by Maurin *et al.* [7] should rather result in a number of discrete reflections, quite close to each other or partially overlapping, but local inhomogeneities of carbon distribution in crystals may be one of the possible reasons of the broadening of reflections. The description and results of the single crystal experiments are given in Table II.

As it was already mentioned, the carbon content at the B site for $Mg(B_{1-x}C_x)_2$ was estimated from the *a* lattice parameter according to Ref. 10. Boron and carbon contents were fixed throughout the entire refinement. The laser ablation ICP mass spectrometry analysis showed no silicon in the substituted crystals, prepared with SiC as a source of carbon. The refinement of Si content in the second crystal converged to zero. Thus, it was assumed, that silicon was not present in crystals and was not considered in the final refinement.

First, $Mg(B_{1-x}C_x)_2$ structures were refined in isotropic approximation with *R*-factors increasing from 0.028 to 0.052 for increasing *x*. Isotropic extinction refinement reduced *R*-factors by 0.004-0.01 with significance level of extinction parameters not less than 3σ (σ - standard deviation). Successively, magnesium occupancy, anisotropic atomic displacement parameters (ADP) and anharmonic ADP were added to the refinements. The results of each step are given in Table III. The refinement of pure $MgB_2$ converged to low *R*-factors without any peculiarities. The refinement of the Mg occupancy differed from unity within 1.5σ and was fixed to unity in the final refinement. It is not the case for the substituted crystals. *R*-factors and residual peaks at difference Fourier maps were significantly higher at each step compared to that of the unsubstituted phase, which means that there are some local distortions compared to the parent structure. The refinement of anisotropic or anharmonic ADP did not improve residual factors significantly or no improvement was observed. In the case of refinement of the anharmonic ADP the probability density function (pdf) for Mg produced strong negative regions and this approximation was rejected in the final refinement. Finally the Mg occupancy was refined for both substituted phases. It resulted in significant improvement of *R*-factors and the 1.5-2 times decrease of residual peaks at the difference Fourier maps. Positional and atomic displacement parameters for each structure are given in Table IV.

The refinement showed Mg deficiency, increasing with increasing carbon content from the 0% for unsubstituted samples to about 10% for the C-substituted samples with the highest carbon amount of $x = 0.15$. The attempt of a direct determination of the ratio of Mg content to B and C content in $Mg(B_{1-x}C_x)_2$ single crystals by electron probe microanalysis (EPMA) was not successful because of difficulties with proper determination of carbon content. Nevertheless, incomplete EPMA data indicate much smaller if any magnesium deficiency as compared to that estimated by x-ray refinement.

A large increase of anisotropic ADP for all atoms with increasing *x* was observed. This fact suggests static disorder for both sites. Such disorder should lead to a shortening of Mg–C distances. In the structure of $MgB_2C_2$ (Ref. 31) Mg–C distances vary in the range 2.25-2.52 Å and similar variations may be expected in the substituted phases as well. It may be the reason for underestimation of magnesium occupancy, because Mg atom may be shifted from ideal position to create asymmetric coordination in unit cells where carbon atoms are present. Small positive residual peaks were observed 0.6-0.8 Å from Mg, and about 1.8-2.3 Å from B(C) site, which corresponds to interatomic distances in the structure mentioned above. Besides, in spite of careful experiments and refinements, *R*-factors for substituted phases remained quite high compared to that of pure $MgB_2$. Thus, it may be suggested that local disorder is present in structures of substituted crystals, which cannot be described in conventional terms.

### C. Superconducting properties

Field cooled magnetization of carbon substituted $MgB_2$ was found to be significantly lower in comparison with that one of unsubstituted compound recorded at the same magnetic field (see Fig. 3). It provides an indication on the increased pinning, most likely due to the local disorder introduced by the carbon substitution, in agreement with the results of x-ray investigations. The small value of the ratio of field-cooled to zero-field-cooled magnetization accompanied with narrow transition width may also indicate on the inhomogenous carbon distribution on the length scale of the distribution of carbon inhomogeneities below the coherence length $\xi$ value rather than on the microscopic phase separation. If C was inhomogeneously distributed on a length scale larger than $\xi$, the magnetization versus temperature should show several steps or a broad transition.

Figure 4 shows the normalized in-plane resistance versus temperature for the $Mg(B_{1-x}C_x)_2$ single crystals with $x = 0$, 0.05 and 0.095. For the non-substituted sample, "the residual resistance ratio" $rrr \cong 7$ (defined as R(300K/R(40K)) is similar or larger than reported in literature.[16, 28, 32, 33] This indicates a good quality of the single crystals. The non-linear increase of the normalized resistance $R(40 K)/R(300 K)$ with carbon substitution is observed and this may result in a saturation effect. Such a saturation of $R(40)/R(300)$ suggests that some changes in the transport properties of the substituted $MgB_2$ single crystals can not be explained by any simple model considering the reduction in the cross-section area of the sample as a main reason of the observed effects.[16] Most likely, the carbon substitution results in creation of more microscopic than macroscopic defects acting as electron scattering centers. This is consistent with the HRTEM results, shown in Fig. 5, where no carbon agglomerates or other extended structure defects have been observed. The appearance of microscopic scattering centers should result in a shorter mean free path and, as a consequence, in decreasing of the superconducting coherence length $\xi$. This should increase $H_{c2}$, as being actually observed for the C-substituted crystals.

The in-plane resistivity $\rho_{ab}$ for the substituted $Mg(B_{1-x}C_x)_2$ single crystal with $x = 0.05$ is shown in Fig. 6. Despite of low $rrr \cong 2$, the crystal exhibits relatively low $\rho_{ab}(300 K) = 20$ μΩcm if compared even with metallic copper ($\rho(300 K) \cong 1.5$ μΩcm, $rrr \cong 20 - 2000$). This feature seems to confirm the rather unusual transport properties of $MgB_2$, as discussed by Rowell.[16] For the sample with $x = 0.05$, $\rho_{ab}$ in the magnetic field $H$ perpendicular to the *ab*-plane is presented in the inset of Fig. 6. A sharp transition to the superconducting state is observed at zero field ($\Delta T_c = 0.05$ K) and this transition widens in higher fields. For $\mu_o H = 2$ and 3 T, a two-step-like behavior in $\rho_{ab}(T)$ is observed at $T_c$, where a sharp drop at lower temperatures may be explained as a result of a vortex melting effect. The shape and temperature of the superconducting transition do not change significantly when the current density $i$ = 2.5 A/cm$^2$ decreases by a factor of 10 indicating more thermally than current induced vortex dynamics. At higher current densities, the vortex driven state is observed with a peak effect that develops at moderate fields. Similar effects have been reported for the non-substituted MgB$_2$ single crystals.[26] The differences in the resistivity of the single crystals with $x$ = 0.05 and 0.095 and for two field configurations ($H$ perpendicular or parallel to the $ab$-plane) seems to be caused mainly by the differences in the pinning force only. Generally, much sharper transitions are observed for the sample with $x$ = 0.095 and for the configuration where $H$ is parallel to the $ab$-plane. Detailed studies of the vortex dynamics in the driven state are performed and will be published separately.

The upper critical fields are derived from the transport measurements, applying two definitions of the transition temperature. Firstly, $T_{on}$ is determined by the linear extrapolation of $R(T)$ to the normal-state resistance from the first sudden drop observed at the transition. Secondly, $T_c$ is defined as a temperature of the linear extrapolation of $R(T)$ (from the similar transition area) to the zero resistance state. This definition of $T_c$ corresponds to a temperature where $R$ drops to zero for a current density small enough to significantly reduce or, depending on $H$, even avoid effects related to the vortex driven state. Thus, $T_{on}$ and $T_c$ are the temperatures corresponding to the upper critical field and the irreversibility field, respectively.[23, 34] However, for $H$ perpendicular to the $ab$-plane and in a case when those critical temperatures differ much one from another, $T_{on}$ seems to be influenced by surface effects and, therefore, $H_{c2}$ is determined more adequate if being associated with $T_c$.[26] This definition of $H_{c2}$ is consistent with bulk measurements of the magnetization,[2,26] specific heat,[27] and thermal conductivity (Ref. 35).

The upper critical fields of the Mg(B$_{1-x}$C$_x$)$_2$ single crystals are shown in Fig. 7 for $H$ parallel ($H_{c2}^{//ab}$) and perpendicular ($H_{c2}^{//c}$) to the $ab$-plane. At 20 K, an anisotropy $\gamma$ = ($H_{c2}^{//ab}/H_{c2}^{//c}$) = 3.4 and 2.8 is measured for the crystals with $x$ = 0.05 and 0.095, respectively. The observed decrease of the anisotropy from $\gamma$(20 K) $\cong$ 4.2-4.5, measured in the non-substituted MgB$_2$[26, 28, 32, 36] to $\gamma$(20 K) $\cong$ 2.8 reflects the general reduction of the electronic anisotropy due to substitution. A small anisotropy $\gamma$(20 K) $\cong$ 2.5 has been reported for a Mg(B$_{1-x}$C$_x$)$_2$ single crystal with $x$ = 0.02, where $H_{c2}$ was measured by local ac susceptibility.[24] It is difficult to comment this result because a large uncertainty in the determination of the carbon content appears in single crystals grown under different conditions. It is known that the real carbon content differs much from the nominal content, as shown by neutron diffraction experiments.[10]

Superior properties of the single crystal with $x$ = 0.095 manifest themselves at temperatures below 10 K (see Fig. 7). At these temperatures, $H_{c2}^{//c}$ for this crystal is larger than $H_{c2}^{//c}$ derived for the less substituted sample with $x$ = 0.05. For $H$ parallel to the $ab$-plane, crossing of the $H_{c2}^{//ab}(T)$ dependences obtained for both single crystals is predicted at 10 K by linear extrapolation of $H_{c2}^{//ab}(T)$ from 20 and 17 K for the crystals with $x$ = 0.05 and 0.095, respectively. The extrapolated $H_{c2}^{//ab}$(10 K) is equal to about 24 T, that is the maximum "zero resistance" $H_{c2}^{//ab}$ (or irreversibility field) observed for any substituted MgB$_2$ single crystals.[23, 24, 25] An important feature revealed in our C-substituted samples is the small difference between the "zero resistance"($H_{c2}^{//c}(T_c)$) and the "onset" ($H_{c2}^{//c}(T_{on})$) upper critical fields. This is shown in Fig. 7, where the upper critical fields, obtained by using both criteria, are plotted. The only exception is observed for the x=0.05 single crystal measured for *H* perpendicular to the *ab*-plane. For this case, a significant difference between $H_{c2}^{//c}(T_c)$ and $H_{c2}^{//c}(T_{on})$ is observed for fields above 1 T. It is suggested that surface effects[27] are the origin, however we are not able to prove this now.

Torque measurements performed as a function of angle in constant magnetic field of 2-3 Tesla at 2.2 K on carbon substituted MgB$_2$ show almost reversible angular dependence (see Fig. 8(a)), similar to the one observed in much lower fields for unsubstituted material.[37] However, for higher magnetic fields irreversibility (in the angular dependence of torque) appears and a significant increase of the torque width is observed (see Fig. 8(b)), with the amplitude of almost one order of magnitude higher than the reversible torque value. The data presented in the Fig. 8(b) indicate the existence of strong pinning in the material, being effective up to the highest magnetic fields of 9 Tesla in our experimental set-up. Importantly, the appearance of strong pinning in Mg(B$_{1-x}$C$_x$)$_2$ single crystals is correlated with the significant shift of the upper critical field and of the irreversibility line towards higher fields/temperatures in the *H-T* phase diagram.[38] The vanishing of the difference between torque signal for clockwise and counterclockwise change of magnetic field presented in Fig. 9 determines the irreversibility field $H_{irr}(\theta)$, which can be considered as estimation for the lower limit of $H_{c2}(\theta)$. Since the difference does not vanish in the full angular range for the field as high as 7 or even 8 Tesla at 2.2 K it means that $H_{c2}^{//c}(2.2 K)$ is higher than 8 T, consistent with the resistance studies (Fig. 7). The last value is more than twice the value of 3.1 T determined for $H_{c2}^{//c}(0)$ of the non-substituted MgB$_2$ single crystals.[2]

A microscopic understanding of the mechanism responsible for the significant increase of $H_{c2}$ in Mg(B$_{1-x}$C$_x$)$_2$ single crystals is desirable. The value of the in-plane Ginzburg-Landau coherence length $\xi_{GL}$ for unsubstituted defect-free MgB$_2$ compound is about 10.2 nm (Ref. 2, 39). Assuming a decrease of the mean free path *l* as a result of the carbon substitution in MgB$_2$ one may expect the decrease of the coherence length and the increase of $H_{c2}$. The estimation of the value of mean free path *l* associated with the observed changes in $H_{c2}$ can be done based on the Pippard relation $1/\xi = 1/\xi_{GL} + 1/l$. With the in-plane coherence length $\xi_{GL}$ of 10.2 nm for unsubstituted compound[39] and estimated the in-plane coherence length $\xi \cong 6.4$ nm, corresponding to $H_{c2}^{//c} \cong 8$ T for Mg(B$_{0.937}$C$_{0.063}$)$_2$, an in-plane value of $l \cong 17.5$ nm can be derived. Periodic in-plane defect distribution with the length scale of 17.5 nm would lead to the most effective pinning in the fields *H*∥*c*-axis as high as 7.8 T. The maximum of the peak effect (PE) for the field configuration *H*∥*c*-axis is observed in a somewhat smaller field of about 6 T. Furthermore, the ratio of the field at the peak effect maximum to $H_{c2}$ is equal to about 0.75 for Mg(B$_{0.937}$C$_{0.063}$)$_2$, lower than that obtained for unsubstituted MgB$_2$, evaluated to be 0.85.

In a previous paper concerning the peak effect in MgB$_2$ we demonstrated that it signifies a disorder-induced phase transition of vortex matter.[38] In such a scenario, the considerably lower reduced field $H/H_{c2}$ at the maximum of the PE at low temperatures in Mg(B$_{0.937}$C$_{0.063}$)$_2$ indicates, that the amount of random distributed point-like disorders is increased by carbon substitution,[40] in agreement with x-ray data and transport results. However, most likely the "extrinsic" scenario alone could not fully describe the significant increase of $H_{c2}$ in carbon substituted $MgB_2$. The impact of carbon substitution on the band structure of $Mg(B_{1-x}C_x)_2$ must be taken into account in the description of the changes observed in the position of $H_{c2}$ in the $H$-$T$ phase diagram as well. Carbon substitution dopes the crystals with additional electrons, increases the Fermi energy $E_F$. However, additional investigations of a possible lower occupation of Mg in $Mg(B_{1-x}C_x)_2$ are necessary to make a definite statement about the shift of $E_F$. A more detailed analysis of the influence of carbon substitution on the electronic structure of $MgB_2$ is presented elsewhere (Ref. 41).

## IV. CONCLUSIONS

The superconducting transition temperature of the carbon substituted $MgB_2$ single crystals can be tuned in a wide temperature range between 10 and 39 K by adjustment of the nominal composition. In the case when SiC is used in the precursor, Si has been found to not incorporate into the structure, revealing C as more favorable dopant. Single crystal structural investigations of $Mg(B_{1-x}C_x)_2$ show elongation of reflections in $c^*$ direction, indicating inhomogeneities of carbon distribution. The length scale of carbon inhomogeneities should be smaller than $\xi$. Torque and transport investigations reveal strong pinning at least up to 9 T for the field $H$ perpendicular to the $ab$-plane. This leads to an enhancement of the critical current and may cause the observed increase of the irreversibility field in $Mg(B_{1-x}C_x)_2$ despite a modest decrease of $T_c$. The observed increase of the upper critical field and the reduction of anisotropy, reflect the subtle interplay between the various microscopic parameters describing the scattering and coupling within and between σ and π bands in pure and substituted $MgB_2$. Intrinsic changes should be considered such as modification of the Fermi surface, σ and π band gaps, intraband and interband scattering.

## Acknowledgements

We are acknowledged to P. Geiser for the help with laser ablation ICP mass spectroscopy investigations and to T. Wenzel, B. Birajdar, and O. Eibl for EPMA data. This work was supported in part by the European Community (program ICA1-CT-2000-70018, Centre of Excellence CELDIS), Swiss Office BBW Nr. 02.0362, Swiss National Foundation, and INTAS (grant No. 99-1136).

TABLE I. Starting composition, estimated carbon content $x$, lattice parameters, and $T_c$ of selected studied single crystals of $Mg(B_{1-x}C_x)_2$.

| Starting composition | Estimated carbon content, $x$ | $a$, Å | $c$, Å | $c/a$ | $T_c$, K |
|---|---|---|---|---|---|
| Without C or SiC substitution | 0 | 3.0849(2) | 3.5187(2) | 1.1406 | 38.5 |
| 3% of B substituted by C | 0.052 | 3.0683(6) | 3.521(1) | 1.1479 | 35.0 |
| 4% of B substituted by C | 0.05 | 3.0687(7) | 3.5217(8) | 1.1476 | 34.3 |
| 5% of B substituted by C | 0.063 | 3.0647(6) | 3.522(2) | 1.1492 | 33.3 |
| 6% of B substituted by C | 0.080 | 3.0592(4) | 3.5208(8) | 1.1509 | 33.0 |
| 7.5% of B substituted by C | 0.087 | 3.0569(5) | 3.5204(9) | 1.1516 | 30.8 |
| 8% of B substituted by C | 0.095 | 3.0544(9) | 3.518(3) | 1.1518 | 30.1 |
| 10% of B substituted by C | 0.100 | 3.0529(9) | 3.5164(2) | 1.1518 | 28.0 |
| 15% of B substituted by SiC | 0.104 | 3.0517(4) | 3.5205(5) | 1.1536 | 25.5 |
| 20% of B substituted by C | 0.150 | 3.0369(5) | 3.5182(5) | 1.1585 | 10.2 |

TABLE II. Summary of crystallographic information for $Mg(B_{1-x}C_x)_2$; $x = 0$, 0.104, and 0.150.

| | | | |
|---|---|---|---|
| Estimated chemical formula – x-ray refinement | $MgB_2$ | $Mg_{0.92}(B_{0.896}C_{0.104})_2$ | $Mg_{0.9}(B_{0.850}C_{0.150})_2$ |
| Crystal system | Hexagonal | | |
| Space group | P6/*mmm* | | |
| Cell constants $a$, $c$ (Å) | 3.0849(2) <br> 3.5187(2) | 3.0517(4) <br> 3.5205(5) | 3.0369(5) <br> 3.5182(5) |
| Volume (Å$^3$) | 28.998 | 28.387 | 28.098 |
| Z | 1 | | |
| $D_{calc}$, g·cm$^{-3}$ | 2.629 | 2.587 | 2.578 |
| Radiation/ Wavelenghth (Å) | MoKα / 0.71073 | | |
| θ range (°) for cell determination | 14.7 - 29.8 | | |
| Linear absorption coefficient (cm$^{-1}$) | 6.15 | 5.85 | 5.75 |
| Temperature (K) | 293 | | |
| Crystal shape | Plate | | |
| Crystal size (mm) | 0.25×0.12×0.07 | 0.24×0.13×0.085 | 0.40×0.16×0.06 |
| Colour | Black | | |
| Diffractometer | CAD4, Graphite monochromator | | |
| Data collection method | ω-θ | ω-θ | ω |
| Absorption correction | Analytical (crystal shape) | | |
| No. of measured reflections | 1230 | 1199 | 791 |
| No. of observed independent reflections | 128 | 124 | 89 |
| Criteria for observed reflections | I > 3σ(I) | | |
| Data collection θ limits (°) | 2 - 70 | 2 – 70 | 2 - 55 |
| Range of $h$, $k$, $l$ | 0→$h$→8; −8→$k$→8; −9→$l$→9 | | 0→$h$→6; −6→$k$→6; −8→$l$→8 |
| $R_{int}$ | 0.025 | 0.023 | 0.024 |
| Refinement on | F | | |
| Extinction | Anisotropic, Becker-Coppens, Type I, Lorenzian distribution | | |
| $R/R_w$ (I > 3σ(I)) | 0.011/0.015 | 0.033/0.044 | 0.036/0.046 |
| Goodness of fit | 1.02 | 3.13 | 2.86 |
| No. of reflections used in refinement | 126 | 122 | 86 |
| No. of refined parameters | 9 | 7 | 7 |
| Weighting scheme | $w = (\sigma^2(F) + (0.012F)^2)^{-1}$ | | |
| $(\Delta/\sigma)_{max}$ | 0.001 | | |
| $\Delta r_{max}$ (e/Å$^{-3}$) +/− | +0.19/-0.18 | +0.64/-0.41 | +0.91/-0.76 |

TABLE III. The comparison of the results for the structure refinement of Mg($B_{1-x}C_x$)$_2$ at different stages.

| No | Parameter | Mg($B_{1-x}C_x$)$_2$ | $R$ | $R_w$ | Goodness of fit | Residual peaks e/Å$^3$ (nearest atom) (positive/negative) | Remarks |
|---|---|---|---|---|---|---|---|
| 1 | Isotropic ADP | 0 | 0.028 | 0.032 | 2.13 | 0.54(Mg)/1.00(Mg) | |
|   |   | 0.104 | 0.044 | 0.061 | 4.26 | 1.21(B)/1.21(Mg) | |
|   |   | 0.150 | 0.052 | 0.069 | 4.2 | 1.33(B)/0.88(Mg) | |
| 2 | Isotropic extinction (Lorenz distribution) | 0 | 0.018 | 0.0025 | 1.71 | 0.68(B)/0.50(Mg) | g*=0.98(15) |
|   |   | 0.104 | 0.038 | 0.057 | 3.98 | 1.47(B)/0.63(Mg) | g=1.0(3) |
|   |   | 0.150 | 0.048 | 0.064 | 3.90 | 1.59(B)/0.53(Mg) | g=1.4(5) |
| 3 | Anisotropic ADP | 0 | 0.022 | 0.024 | 1.65 | 0.46(Mg)/0.61(Mg) | |
|   |   | 0.104 | 0.43 | 0.059 | 4.13 | 1.23(B)/1.05(Mg) | |
|   |   | 0.150 | 0.053 | 0.065 | 3.98 | 1.44(Mg)/0.94(Mg) | |
| 4 | Isotropic extinction and Anisotropic ADP | 0 | 0.011 | 0.016 | 1.07 | 0.20(B)/0.19(Mg) | g=0.97(9) |
|   |   | 0.104 | 0.037 | 0.054 | 3.84 | 1.46(B)/0.60(Mg) | g=1.0(3) |
|   |   | 0.150 | 0.050 | 0.059 | 3.64 | 1.61(B)/0.93(Mg) | g=1.4(4) |
| 5 | No.4 + Mg occupation | 0 | 0.012 | 0.016 | 1.06 | 0.20(Mg)/0.19(Mg) | Mg=0.994(4) |
|   |   | 0.104 | 0.033 | 0.44 | 3.13 | 0.64(B)/0.41(Mg) | Mg=0.910(11) |
|   |   | 0.150 | 0.036 | 0.046 | 2.86 | 0.91(B)/0.76(Mg) | Mg=0.891(14) |
| 6 | No.4 +Mg anharmonic ADP 4 terms | 0 | 0.011 | 0.015 | 1.02 | 0.19(Mg)/0.18(Mg) | +/- pdf = 100/0.48 % |
|   |   | 0.104 | 0.034 | 0.052 | 3.72 | 1.38(B)/0.61(Mg) | +/- pdf = 100/205 % |
|   |   | 0.150** | 0.046 | 0.058 | 3.59 | 1.49(Mg)/0.80(Mg) | +/- pdf = 100/183 % |

\* Extinction parameter

\*\* Refined without extinction; extinction parameter was less than 2σ.

TABLE IV. Positional and atomic displacement (Å$^2$) parameters and Mg-B(C) bond length for Mg(B$_{1-x}$C$_x$)$_2$ with $x$ = 0, 0.104, 0.150.

| Parameter | MgB$_2$ | Mg$_{0.91}$(B$_{0.896}$C$_{0.104}$)$_2$ | Mg$_{0.9}$(B$_{0.850}$C$_{0.150}$)$_2$ |
|---|---|---|---|
| *Occupation,* Mg | 1 | 0.910(11) | 0.891(14) |
| $x$, Mg | 0 | | |
| $y$, Mg | 0 | | |
| $z$, Mg | 0 | | |
| $x$, B/C | 1/3 | | |
| $y$, B/C | 2/3 | | |
| $z$, B/C | ½ | | |
| $U_{11}\times10^3$, Mg | 4.95(17) | 8.54(18) | 11.3(3) |
| $U_{33}\times10^3$, Mg | 5.81(18) | 9.05(19) | 12.4(4) |
| $U_{12}\times10^3$, Mg | ($U_{11}\times10^3$)/2 | ($U_{11}\times10^3$)/2 | ($U_{11}\times10^3$)/2 |
| $U_{11}\times10^3$, B/C | 4.29(10) | 9.8(3) | 13.4(5) |
| $U_{33}\times10^3$, B/C | 5.79(11) | 9.2(3) | 13.8(6) |
| $U_{12}\times10^3$, B/C | ($U_{11}\times10^3$)/2 | ($U_{11}\times10^3$)/2 | ($U_{11}\times10^3$)/2 |
| Mg-B bond length, Å | 2.503 | 2.491 | 2.483 |
| B-B bond length, Å | 1.781 | 1.762 | 1.752 |

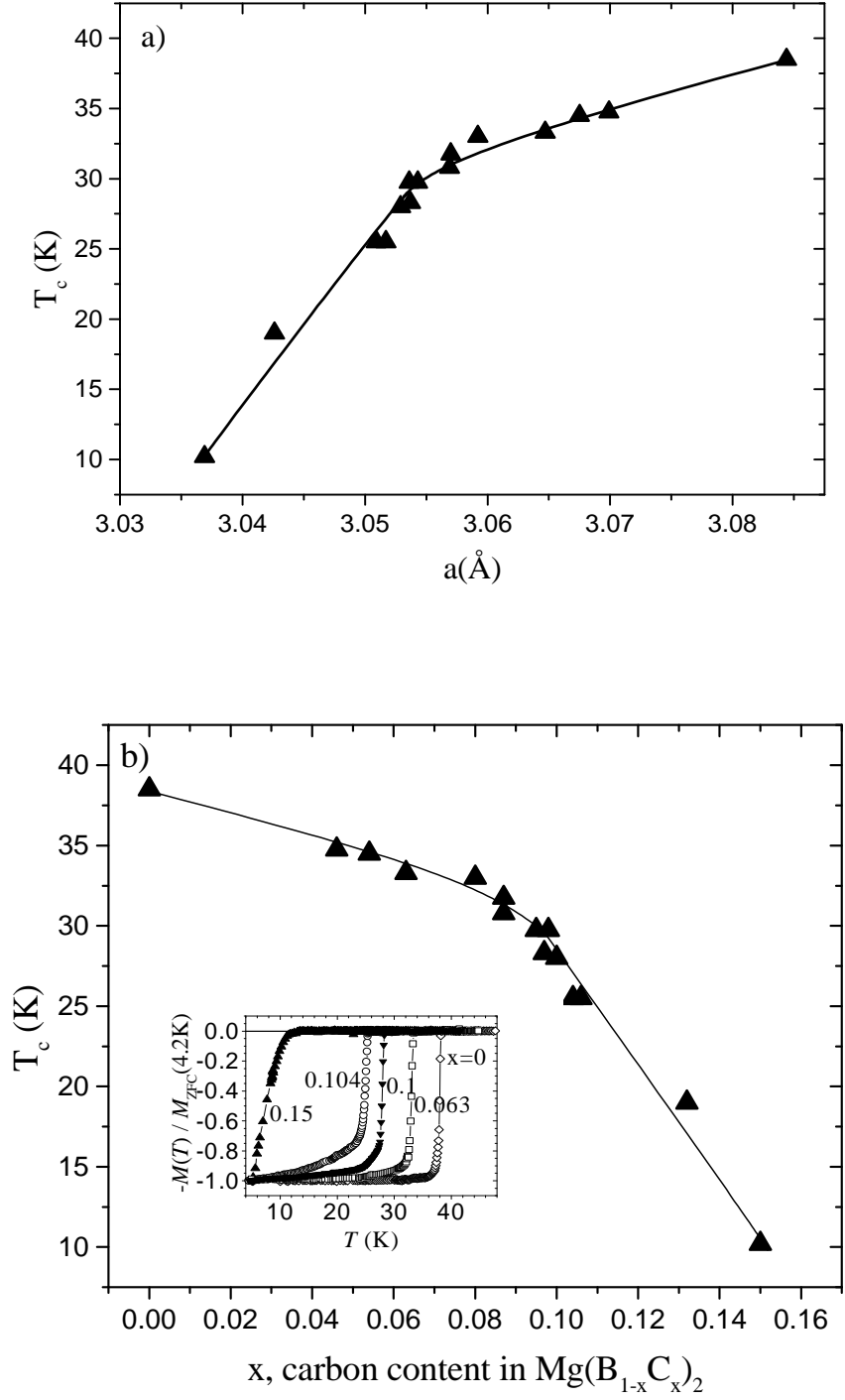

FIG. 1. Variation of $T_c$ as a function of lattice parameter $a$ (a) and carbon concentration (b). Inset in Fig. 1(b) shows magnetization curves of $(MgB_{1-x}C_x)_2$ crystals with various carbon content.

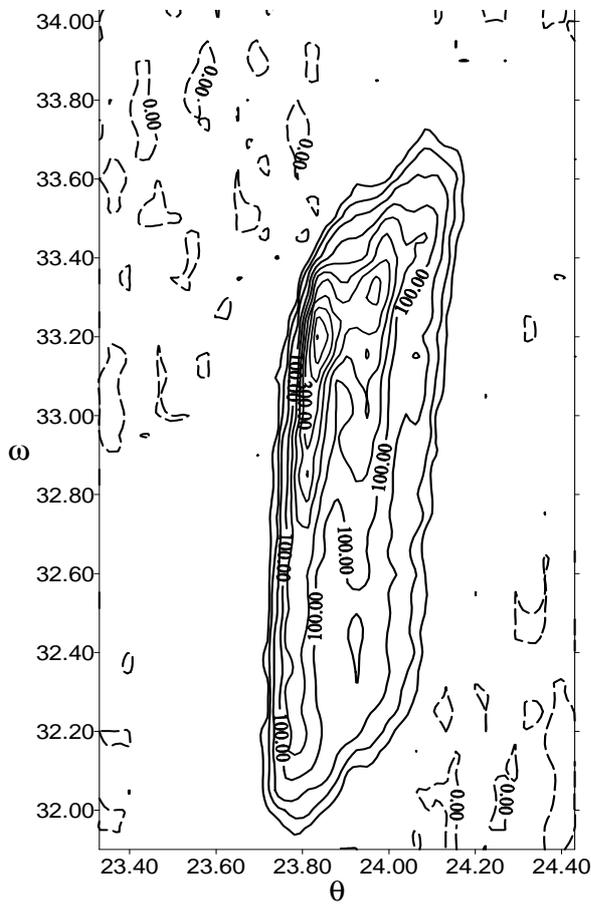

FIG. 2(a). ω-θ scan of 300 reflection of Mg(B$_{0.85}$C$_{0.15}$)$_2$; $c^*$ is parallel to the plane and to ω axis, $a^*b^*$ is perpendicular to the plane; contour at 0 (dashed line), 12.5, 25, 50, 100, then contour step 100 sec$^{-1}$. The reflection of the C substituted MgB$_2$ crystal shows an elongation in $c^*$ direction.

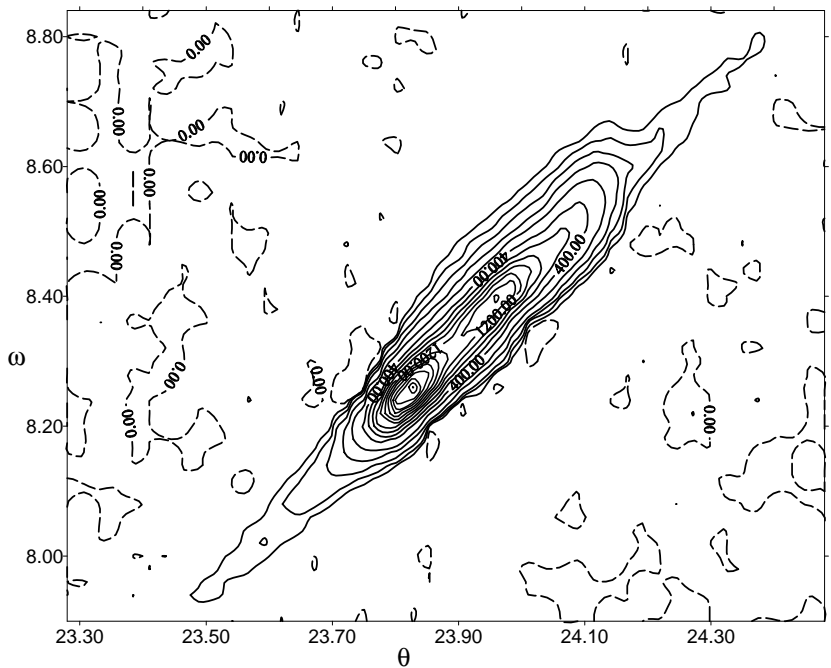

FIG. 2(b). ω-θ scan of 300 reflection of Mg(B$_{0.85}$C$_{0.15}$)$_2$; $c^*$ is perpendicular to the plane, $a^*b^*$ is parallel to the plane, $a^*$ is parallel to the reflection; contour at 0 (dashed line), 12.5, 25, 50, 100, 200, then contour step 200 sec$^{-1}$. The shape of the reflection is similar to the unsubstituted MgB$_2$ crystal. Two peaks correspond to K$_{α1}$ and K$_{α2}$ reflections.

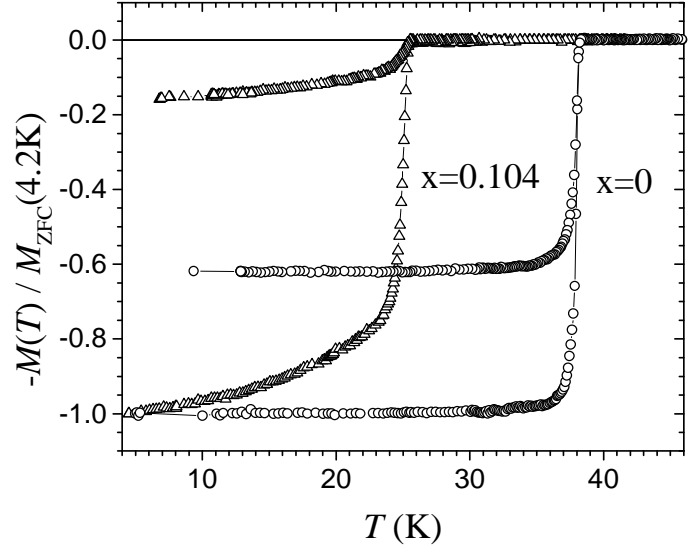

FIG. 3. Zero field cooled and field cooled magnetization of unsubstituted ( x = 0) and carbon substituted (x = 0.104) Mg($B_{1-x}C_x$)$_2$ single crystals.

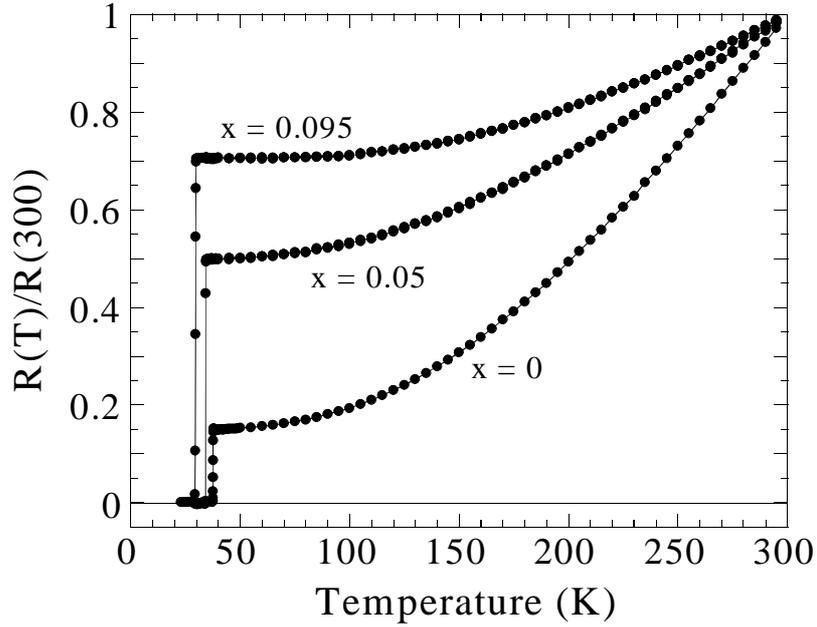

FIG. 4. Temperature dependence of the in-plane resistance ratio $R(T)/R(300\text{ K})$ for Mg($B_{1-x}C_x$)$_2$ single crystals, non-substituted ($x = 0$) and carbon-substituted with $x = 0.05$ and 0.095. The superconducting transition temperature measured with low current density ($i = 2.5$ A/cm$^2$) is $T_c$ = 38.2, 34.3, and 30.1 K for the crystals with $x = 0$, 0.05, and 0.095, respectively.

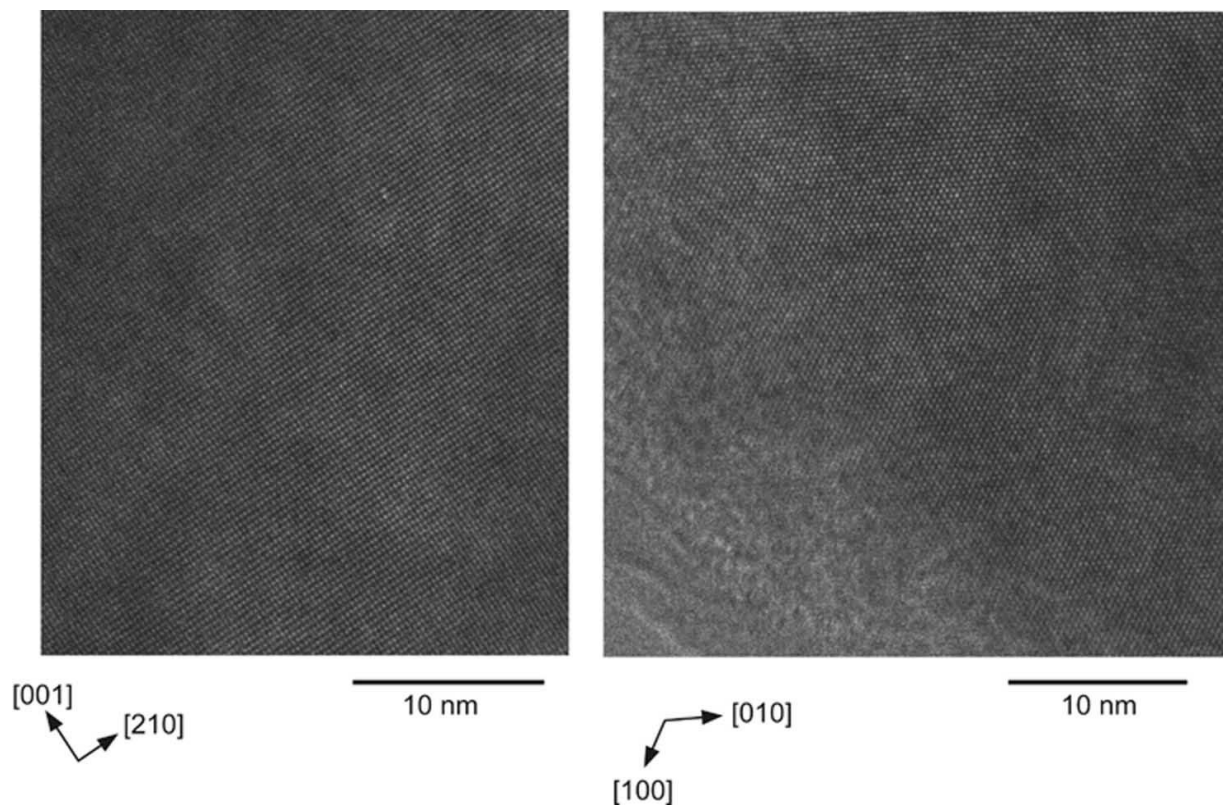

FIG. 5. High-resolution electron microscopy of a Mg(B$_{0.95}$C$_{0.05}$)$_2$ sample (carried out in a Philips CM30 at 300 kV) The images were recorded along the [0 -1 0] direction (left) and along the [001] direction (right). A high level of structural perfection is observed. The low image contrast in the lower left part of the right image is due to an amorphous surface layer of the Mg(B$_{0.95}$C$_{0.05}$)$_2$ grain, which is caused by the sample preparation for measurement.

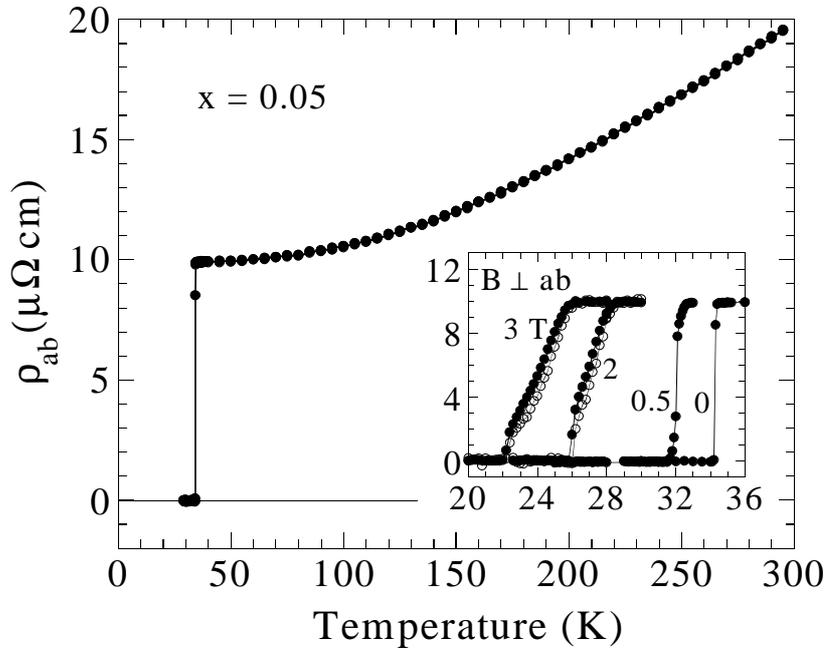

FIG. 6. Temperature dependence of the in-plane resistivity for the Mg(B$_{1-x}$C$_x$)$_2$ single crystal with $x = 0.05$. The inset shows the resistivity measured at several applied fields oriented perpendicular to the *ab*-plane. These measurements were performed at two current densities of $i = 2.5$ A/cm$^2$ (closed circles) and 0.25 A/cm$^2$ (open circles).

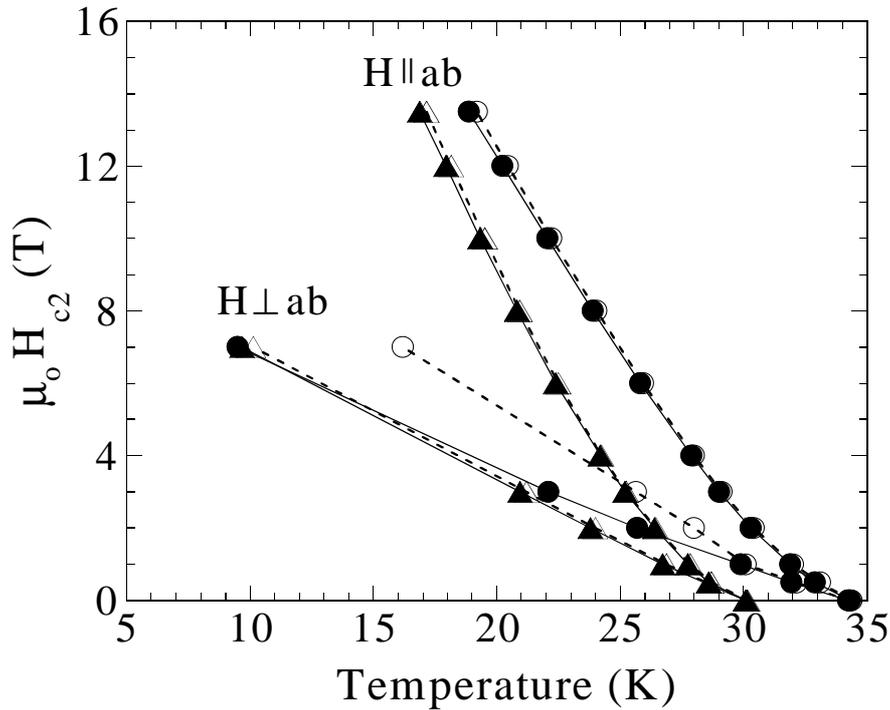

FIG. 7. Magnetic phase diagram of the Mg(B$_{1-x}$C$_x$)$_2$ single crystals with $x = 0.05$ (circles) and 0.095 (triangles). The upper critical fields are derived using the "zero resistance" (closed symbols, solid lines) and the "onset" (open symbols, broken lines) transition temperature definitions.

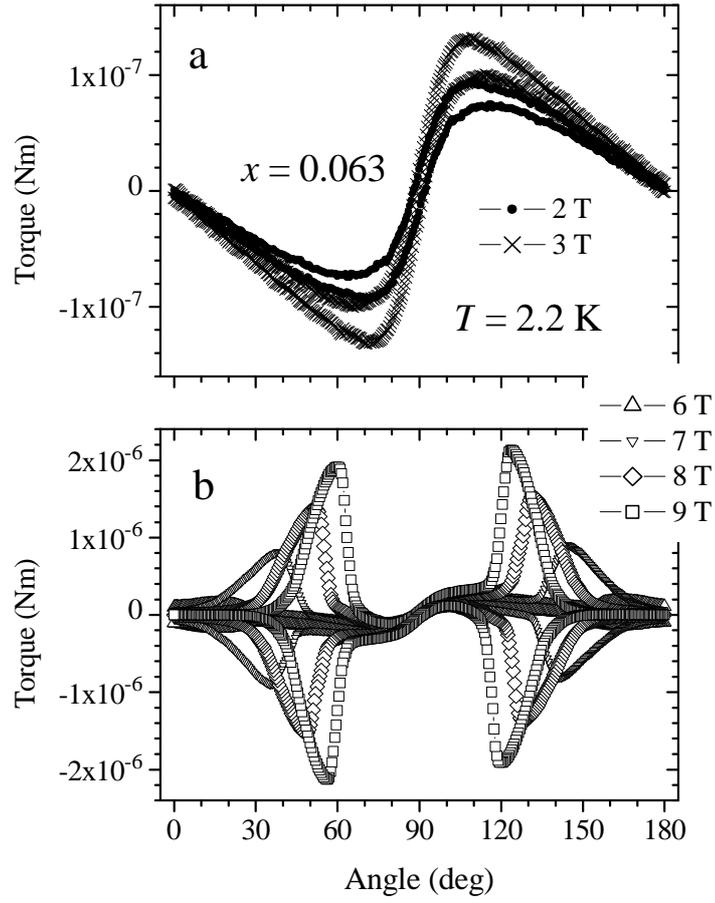

FIG. 8. Angular dependence of the torque for Mg(B$_{0.937}$C$_{0.063}$)$_2$ in different magnetic fields at 2.2 K.

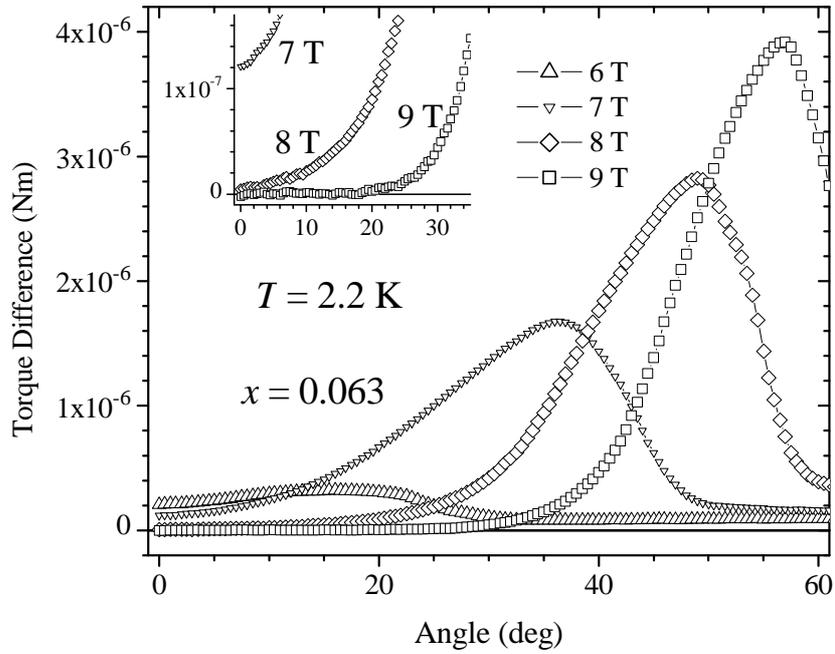

FIG. 9. Difference between torque signal for clockwise and counterclockwise change of the field direction in different magnetic fields for Mg(B$_{0.937}$C$_{0.063}$)$_2$ single crystal at 2.2 K.